\documentclass[aps,prb,twocolumn,showpacs,floatfix,superscriptaddress]{revtex4}
\usepackage{amsfonts}
\usepackage{amsmath}
\usepackage{amssymb}
\usepackage{graphicx}
\usepackage{multirow}
\usepackage{dcolumn}%
\usepackage{tabularx}
\usepackage{footnote}
\usepackage{todonotes}
\usepackage{soul}
\usepackage{float}
\providecommand{\U}[1]{\protect\rule{.1in}{.1in}}
\begin{document}
\preprint{ }
\title{Optical spectrum of bottom-up graphene nanoribbons: towards efficient atom-thick excitonic solar cells}
\author{Cesar E.P. Villegas}
\affiliation{Instituto de F\'{\i}sica Te\'{o}rica, Universidade Estadual Paulista (UNESP),
S\~{a}o Paulo, Brazil.}
\author{P.B. Mendon\c{c}a}
\affiliation{Instituto de F\'{\i}sica, Universidade de S\~{a}o Paulo, CP 66318, 05315-970,
S\~{a}o Paulo, SP, Brazil }
\author{A.R. Rocha}
\affiliation{Instituto de F\'{\i}sica Te\'{o}rica, Universidade Estadual Paulista (UNESP),
S\~{a}o Paulo, Brazil.}
\begin{abstract}
Recently, atomically well-defined cove-shaped graphene nanoribbons have been obtained
using bottom-up synthesis. These nanoribbons have an optical gap in the visible range of
the spectrum which make them candidates for donor materials in photovoltaic devices.
From the atomistic point of view, their
electronic and optical properties are not clearly understood. Therefore, in this work we
carry out \textit{ab-initio} density functional theory calculations
combine with many-body perturbation formalism to study their electronic and optical properties.
Through the comparison with experimental measurements, we show that an accurate description of the nanoribbon's optical properties
requires the inclusion of electron-hole correlation effects. The energy, binding energy and the corresponding excitonic transitions involved
are analyzed.
We found that in contrast to zigzag graphene nanoribbons, the excitonic peaks in the absorption
spectrum are a consequence of a group of transitions involving the first and second conduction and valence
bands. Finally, we estimate some relevant optical properties that strengthen
the potential of these nanoribbons for acting as a donor materials in photovoltaic.
\end{abstract}
\maketitle
\section{Introduction}

The discovery of graphene and its unique electronic properties\cite{rev,rev2}
has led to a number proposals %, which indicate it holds great promise
for novel electronic,\cite{grap,prospec1,prospec} optoelectronic,\cite{gopto} and photovoltaic\cite{gphoto}
devices. For instance, Bernardi {\it et al.}\cite{hete} have recently proposed that graphene combined with transition
metal dichalcogenides could yield high solar energy absorption rates - up to three orders of magnitude higher
than the most efficient solar cells - paving the way for next-generation photovoltaics.
Nevertheless, the lack of a band gap has hindered further development of
graphene-only devices. Particularly in the case of photovoltaic applications, there are a number of key ingredients
which are important in designing
materials for harvesting solar energy. Besides the presence of a gap, which should
preferably coincide with the visible spectrum range,
the presence of Frenkel excitons (strongly bound excitons) is also important.\cite{frenk_photo}

There have been several attempts to induce such a band gap in graphene-like materials, including
adsorbed atoms,\cite{adsorp,adsorp2} strain engineering,\cite{strain,strain2} doping,\cite{doped,doped2} and lateral
confinement.\cite{ribbon,ribbon1,nanorods,nanoroadsJames} In particular, graphene nanoribbons
(GNRs),\cite{ribbon,ribbon1} laterally constrained graphene sheets, exhibit semiconducting
behavior in narrow samples less than ten nanometers wide.\cite{ribbon2} It has also been theoretically shown
that the electronic properties of GNRs are strongly dependent
on the edge geometry and width,\cite{ribbon3} which could be used to tune the band gap.\cite{opto}
%This fact, enables the
%possibility of tuning their energy gap, and, thus enabling them for nano-electronic and optoelectronic\cite{opto} applications.
%Recently, photovoltaic devices based on atomically thin materials such as graphene and transition metal dichalcogenides have shown great potential as
%they could achieve high power densities
%up to three orders of magnitude higher than the most efficient solar cells.\cite{hete}
Thus, using some features of GNRs, such as the semiconducting behavior and their quasi-1D character,
which typically enhances the exciton binding energies,\cite{qp2} one might envision employing them as
long-exciton lifetime donor materials for photovoltaic applications.

However, tailoring GNRs by using top-down approaches, such
as lithography,\cite{lito} unzipping of carbon nanotubes,\cite{unzip} and
sonification patterning\cite{soni} is still a challenging task, since these
methods still lack the atomic precision control for tailoring the
edge structure. Recently, bottom-up chemical
synthesis has led to the accurate patterning of GNRs
with atomically precise edges.\cite{bottom-up} These synthesis methods
generally use small appropriate polyphenylene molecules as precursors that can be
mediated by either solutions\cite{solu,naturecomm,natureche,naturecomm2} or metallic surfaces.\cite{cai}
%For instance, Cai {\it et al.} use surface-mediated approaches to obtain well-defined armchair
%GNRs with lengths up to 60 nm.
More recently, M\"ullen's group reported solution-mediated synthesis of GNRs over 200 nm long
and chemically precise widths of approximately 1 nm.\cite{naturecomm,natureche}

%%%%%%%%%%%%%FIG1%%%%%%%%%%%%%%%%%%%%%%%%%%%%%%%%%
\begin{figure}[t]
%\centering
\includegraphics[width=0.95\columnwidth]{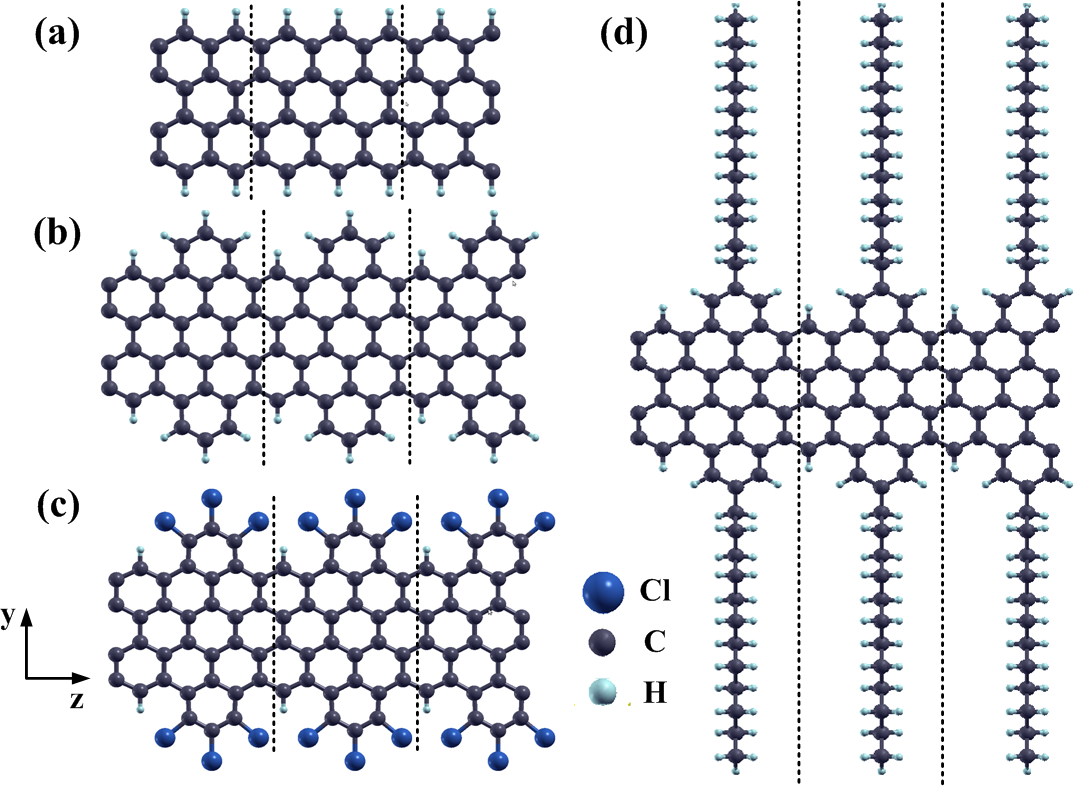}
\caption{\small{Schematic representation of the four graphene nanoribbons
studied. (a) 4-ZGNR, (b)H-CGNR,
(c) Cl-CGNR, and (d) C$_{12}$H$_{25}$-CGNR. The vertical doted lines indicates the
length of the unit cell considered in the simulations. The periodic direction
is along the $z$-axis.
%\textcolor{red}{\st{The structures were generated using the softwareXCrySDen}}\cite{crysden}
}} \label{fig.1}
%{\rule{4cm}{2cm}}
\end{figure}
%%%%%%%%%%%%%FIG1%%%%%%%%%%%%%%%%%%%%%%%%%%%%%%%%%

One of the resultant structures in the aforementioned studies (the so-called GNR 3) is composed of a
4-zigzag GNR with benzo-fused rings on both edges forming a cove-shaped GNR (CGNR). These, subsequently, bond
to either a chain of C$_{12}$H$_{25}$\cite{natureche} or to
Chlorine atoms\cite{naturecomm} (see Fig. \ref{fig.1}c and \ref{fig.1}d). The authors show that, for both structures, the optical gap lie in
the visible part of the
spectrum, ranging from 1.7 to 1.9 eV, thus suggesting that % This result, combined with its nanometer scale, and with the possibility
%of tuning the optical band gap by functionalizing the edges of GNR 3,
%can turn
this type of system could be used as  a promissing donor material for nanoscopic photovoltaic devices.\cite{opto}

In order to better understand the electronic properties of these graphene nanoribbons, we carry
out \textit{ab-initio} density functional theory
(DFT) calculations using the G$_0$W$_0$ correction to better describe the exchange and correlation potential.
Subsequently we included
electron-hole correlation effects via the Bethe-Salpeter equation (BSE) to simulate the optical spectrum.
%\st{ to simulate the optical properties of different types of GNRs}.
We elucidate the most important optical transitions as well as the exciton binding energies, comparing
our results to recent experiments.\cite{naturecomm,natureche}
%performed by Narita {\it et al.}\cite{naturechem}
%\todo[inline]{As figuras da absorbancia são s\'o do Narita?}
%\textcolor{red}{\st{Concerning the electronic structure, the results show that the presence of
%dodecane chains bound to the edges of the benzo-fused rings do not play
%a significant role on the states around the Fermi level.}}
%\st{We also show, through the comparison with experimental measurements, that
%an accurate description of the GNR's optical properties require a level of theory that
%includes many-body correlation effects.}
Most importantly, we estimate the short-circuit
current density for these atomically thin bottom-up GNRs, finding attractive values
that go up to one order of magnitude higher than those estimated for a nanometer-thick
Si and GaAs, two of the foremost systems used in photovoltaics, particularly as donor materials.

\section{Theory and Methodology}

We carry out this study considering four different GNRs whose widths are smaller than 1nm
as shown in Figure \ref{fig.1}. Although a pristine zig-zag nanoribbon has not been realized experimentally, the
4-zigzag GNR (4-ZGNR) in figure \ref{fig.1}a is included as a
reference for comparison purposes. It is considered in the lowest energy configuration,
with antiferromagnetic order between
carbon atoms of opposite edges, which induces a finite energy gap.\cite{loui} The two other
structures (Fig. \ref{fig.1}c and Fig \ref{fig.1}d) correspond to those obtained experimentally, namely a 4-ZGNR with
benzo-fused rings on either side terminated with either Chlorine (Cl-CGNR) atoms or a dodecane
(C$_{12}$H$_{25}$-CGNR) chain. Finally, in order to determine the effects of the side chains on the electronic
structure we also performed calculations on a C$_{12}$H$_{25}$-CGNR saturated only with hydrogen atoms, hereafter called H-CGNR.

The calculations were performed in three steps. First, plane-wave
density functional theory\cite{dft1,dft2} is used to obtain the electronic ground-state by means of
Perdue-Burke-Ernzerhof (PBE)\cite{pbe} exchange-correlation functional currently implemented
in the Quantum Espresso package.\cite{QE} We employed norm-conserving pseudopotentials
and used a 90 Ry kinetic energy cutoff and a k-sampling grid in the Monkhorst-Pack
scheme of 1 $\times$ 1 $\times$ 32. The structures are fully optimized to their
equilibrium position with forces smaller than 0.02 eV/\AA. In all cases a supercell with
a vacuum region of 16 \AA \ in both directions perpendicular to the $z$ axis was used. This
is large enough to avoid spurious interactions between images.

Next, within the G$_{0}$W$_{0}$ approximation, %for the self-energy
the quasiparticle energies are obtained considering the Khom-Sham eigenstates and
eigenvalues as a starting point,
\begin{equation}
E_{n}^{QP}=E_{n}^{KS}+\left\langle \Psi _{n}^{KS}\right\vert
\text{ }\Sigma (E_{n}^{QP})-V_{XC}\left\vert \Psi _{n}^{KS}\right\rangle,
\end{equation}
where $V_{XC}$, is the exchange correlation potential at the DFT level and $\Sigma$
is the self-energy operator. The screened
Coulomb potential W$_{0}$ is calculated within the Plasmon-Pole approach
%with an energy cutoff of 4 Ry and
including 460 unoccupied bands. In addition, we used a truncated
screened Coulomb interaction to avoid image effects between periodic cells.
%, that is a rectangular-shape due to the geometry of the structures.

Finally, the electron-hole
interactions, relevant in photo-excitation processes, are included by solving
the Bethe-Salpeter equation\cite{bse} for each excitonic state $S$
\begin{equation}
\sum_{v' c' \textbf{k'}}\left\langle vc\textbf{k}\right\vert
\text{ }K^{eh}\left\vert v'c'\textbf{k'} \right\rangle+(E_{c\textbf{k}}^{QP}-
E_{v\textbf{k}}^{QP})A_{vc\textbf{k}}^{S}=\Omega^{S}A_{vc\textbf{k}}^{S}
\end{equation}%
where $A_{vc\textbf{k}}^{S}$, $\Omega^{S}$ is the exciton eigenfunction and
eigenvalues for the $S-th$ exciton respectively, and $K^{eh}$ is the electron-hole interaction kernel.
Together with the Tamm-Dancoff approximation,\cite{tammdancoff} eight valence bands and eight
conduction bands are included to solve the BSE.
Once the excitonic eigenvalues and eigenfunctions are obtained,
one can calculate the optical absorption through the imaginary part of the dielectric
function,
\begin{equation}
\epsilon_{2}(\omega)=16\pi^{2}e^{2}/\omega^{2}\sum_{S}\vert \textbf{e}.\left\langle 0\right\vert
\textbf{v}\left\vert S \right\rangle\vert^{2} \delta(\omega-\Omega^{S})
\end{equation}%
where, \textbf{v} corresponds to the velocity operator along the direction
of the polarization of light \textbf{e}, which is chosen parallel along the ribbon axis,
since the significant optical response in 1D systems take place in this direction.\cite{23}
We stress that a finer k-grid sample of 1 $\times$ 1 $\times$ 128 was
used during the BSE procedure. The G$_{0}$W$_{0}$ and BSE
calculations were performed using the BerkeleyGW package.\cite{bgw}

\section{Results}

\subsection{Electronic Structure}
%%%%%%%%%%%%%FIG2%%%%%%%%%%%%%%%%%%%%%%%%%%%%%%%%%
\begin{figure}[h]
\centering
\includegraphics[width=1.05\columnwidth,height=7cm]{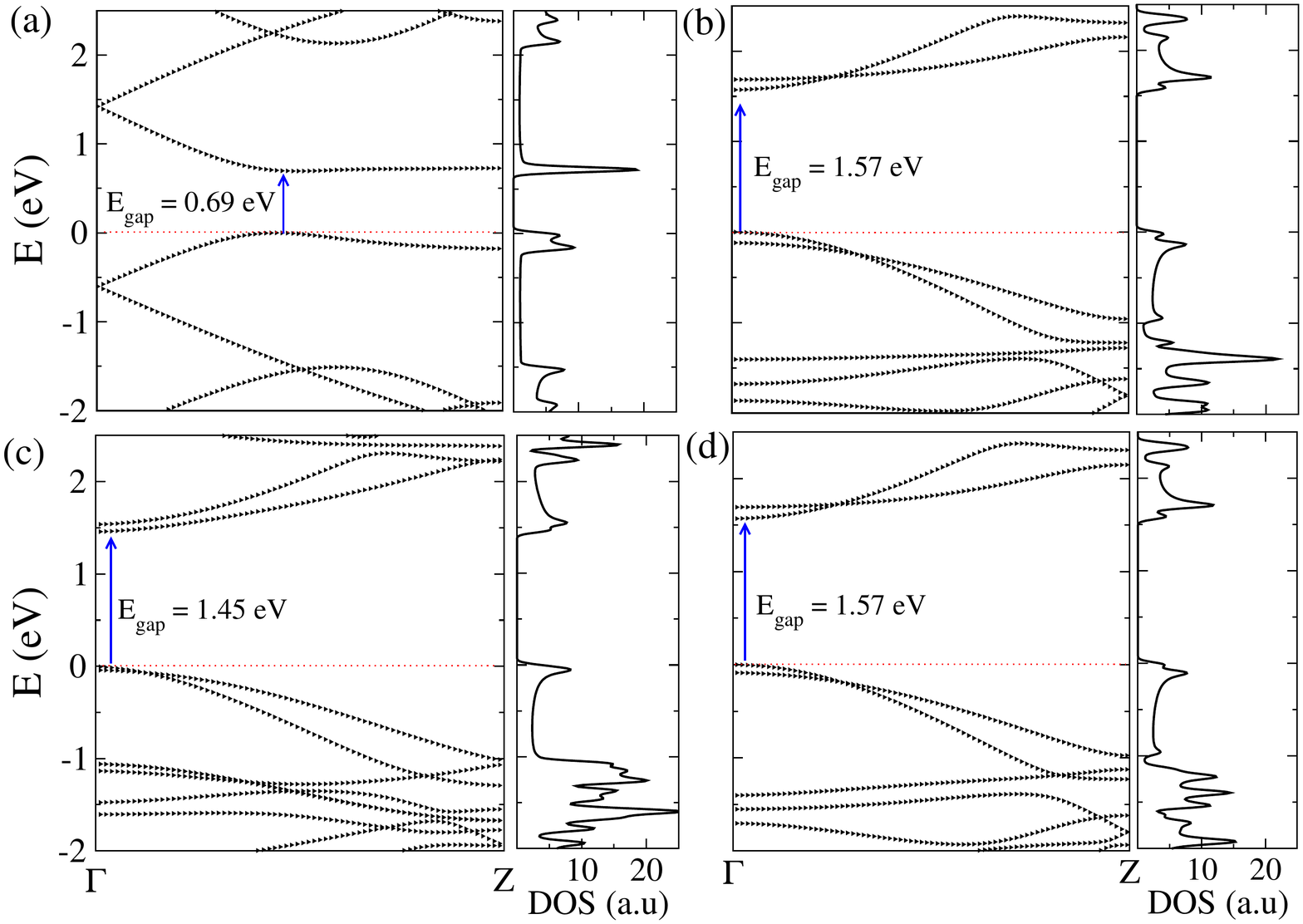}
\caption{\small{Energy bands and its corresponding density
of states calculated with GGA for: (a) 4-ZGNR, (b) H-CGNR, (c) Cl-CGNR,
and (d) C$_{12}$H$_{25}$-CGNR. The (blue) arrow indicates the energy gap. Since all
the nanoribbons are semiconducting, the Fermi level was positioned at the top of the valence band.}} \label{fig.2}
%{\rule{4cm}{2cm}}
\end{figure}
%%%%%%%%%%%%%FIG2%%%%%%%%%%%%%%%%%%%%%%%%%%%%%%%%%

Before going on to discuss the optical properties of the nanoribbons, we initially look at the electronic structure at the GGA level.
It gives an insight to the character of the bands and the most important states contributing to the valence and condution bands. As we will later
see, these will not change with the inclusion of many-body corrections.
Figure \ref{fig.2} shows the electronic band structure for all structures obtained using GGA.
A direct comparison between the
different panels shows the differences on the electronic structure
of the 4-ZGNR (Fig. \ref{fig.2}a)  after the benzo-fused rings are added to the edges (Fig. \ref{fig.2}(b-d)). Firstly, the
valence band maximum (VBM), and conduction band minimum (CBM) are moved
to the $\Gamma$ point, and two of the topmost VBs and the bottommost CBs cross at approximately $k_z=\pi/3a$.
In addition, the band gap is increased by approximately 1 eV and flat states arise in
the energy interval from -1 to -2 eV.

By comparing the three cove-shaped we first notice that the C$_{12}$H$_{25}$ side chains have no significant
effects on the dispersion relation of figure \ref{fig.2}d in the energy window ranging from -1 to 2.5 eV compared to the hydrogen-saturated
case. At the same time, when we substitute the H atoms by Chlorine there is a small reduction in the GGA band gap of
approximately 0.1 eV. Most importantly the presence of Chlorine atoms brings the two valence bands
closer together (making them almost degenerate) at the $\Gamma$ point and inverts the two bottommost conduction bands.

%that  Figs. \ref{fig.2}b and \ref{fig.2}c,
%the situation in which the benzo-fused ring at the edges are bonded to either H or Cl-atoms
%respectively, we note that the curves
%a small reduction in the energy gap of about 0.1 eV is observed.
%Most importantly
%\st{split the crossing %break the energy degeneracy
%observed in the two lowest conduction bands of H-GNR-3, which takes place at
%around a quarter of the energy dispersion.}
%Furthermore, from Fig. 2(d) it is depicted that the presence of the dodecane chains bounded to
%the benzo-fused rings do not induce significant differences
%on the electronic structure of H-GNR-3 within an energy window
%from -1.0 to 2.5 eV as can be seen comparing Figs. 2(b) and 2(d).
%Furthermore we notice - by comparing Figs \ref{fig.2}b and \ref{fig.2}d - that the side chains have no significant
%effects on the dispersion relation in the energy window ranging from -1 to 2.5 eV.

%%%%%%%%%%%%FIG3%%%%%%%%%%%%%%%%%%%%%%%%%%%%%%%%%

\begin{figure}[h]
\centering
\includegraphics[width=0.90\columnwidth]{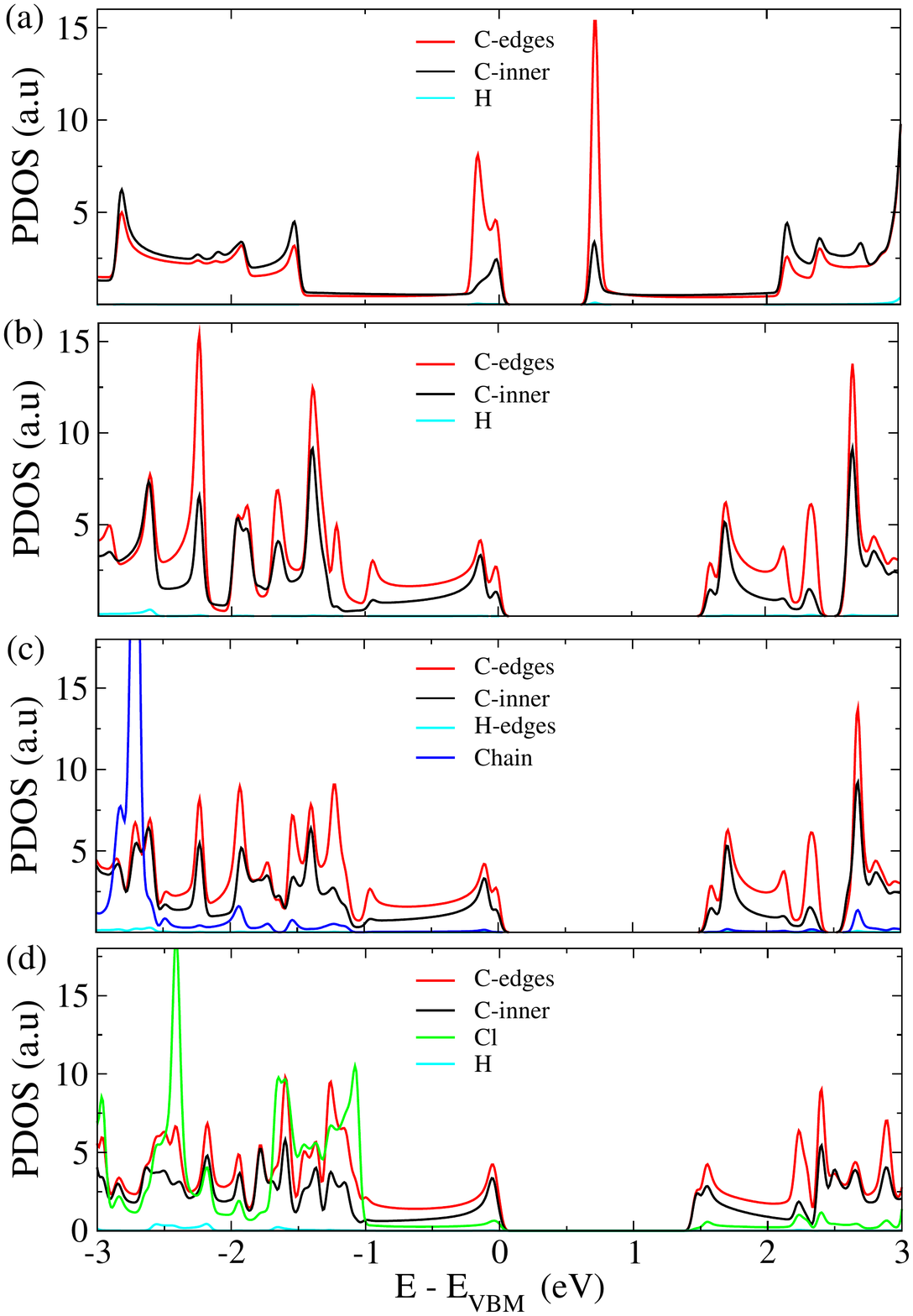}
\caption{\small{Projected density of states for each atomic
species at different spacial regions, calculated with GGA functional. The C-inner and chain states
are related to the atoms located between the edges and those belonging
to the C$_{12}$H$_{25}$ chain, respectively. From top to bottom:
(a) 4-ZGNR, (b) H-CGNR, (c) C$_{12}$H$_{25}$-CGNR, and (d) Cl-CGNR.}} \label{fig.3}
%{\rule{4cm}{2cm}}
\end{figure}
%%%%%%%%%%%%%FIG3%%%%%%%%%%%%%%%%%%%%%%%%%%%%%%%%%

To better understand the role of each atomic species in the electronic structure, we present
the projected density of states (PDOS) for the four GNRs in Fig. \ref{fig.3}. For the case of the 4ZGNR
one notices the presence of localized states both at the top of the
conduction band, and the bottom of the valence band. These can be associated with the well-known
anti-ferromagnetic edges.\cite{afmzgnr} That is not the case for the other nanoribbons, for which there
is a similar balance between carbon states located at the edges and at the inner part.
%\todo[inline]{Could we say that this is similar to armchair ribbons?}
The PDOS also shows that Chlorine states give a contribution to both the CBM and the VBM.
%\todo[inline]{It would be interesting to discern whether the Chlorine character can be associated with specific
%bands. This could give an insight why there is a direct transition here, but not for the H-ZGNR.}
In addition, the C states assigned to the dodecane chain only come into play
for energy values lower than -2.5 eV. This result combined with similarity in the band structure
leads to the conclusion that they play no role in the optical properties of these systems. Thus, hereafter
 our predictions concerning the relevant optical
transitions (those around the Fermi level) are carried out in the absence of the C$_{12}$H$_{25}$ chains.

\subsection{Optical properties}

%This fact put in evidence that excitonic effects in the absorption spectrum plays a key role
%in predicting accurately optical properties.

%%%%%%%%%%%%%%%%%%%%%%%%%%%%%%%%%%%%%%%%%%%%%%5
%\use{multirow}
\begin{table*}[t]
\caption{\small{Energy band gap at different levels of theory (second to third columns).
Fourth and fifth column show the theoretical and experimental values for the absorption
maximum peak. Sixth to eleven-th column stands for the calculated binding energies for different
excitonic states. The transitions where no exciton is found are left blanck. All values are in eV.}\label{table1}}
%title name of the table
%\tiny
%\caption{Performance After Post Filtering}  % title name of the table
%\centering                            % centering table
%\par
\par
\begin{center}
%\setlength{\tabcolsep}{1pt} %\setlength{\extrarowheight}{1.5pt}
%\begin{ruledtabular}
%{\small

\begin{tabular}{lcccccccccccc}%{0.45\textwidth}[c]
\hline\hline
& & &  &  &  &  &  & &  &  \\[-5pt]
&  \multicolumn{4}{c}{ Band gap (eV)} & \multicolumn{6}{c}{Exciton binding energy (eV) } \\[5pt]
 & \ \ GGA & \ \  GW & \ \ BSE & \ \  Exp & \ \  E$_{1}^{11}$ & \ \ E$_{2}^{11}$ & \ \  E$_{3}^{11}$ & \ \  E$_{1}^{21}$ & \ \ E$_{1}^{12}$ & \ \ E$_{1}^{22}$  \\[0.3ex]
\hline
& & &  &  &  &  &  & &  &  \\
\raisebox{0.3ex}{4-ZGNR} &  \raisebox{0.3ex}{0.69} & \raisebox{0.3ex}{2.02} &
 \raisebox{0.3ex}{0.8} &  \raisebox{0.1ex}{-} & \raisebox{0.3ex}{1.22} & \raisebox{0.3ex}{0.87}  & \raisebox{0.3ex}{0.5} & \raisebox{0.3ex}{1.49} & \raisebox{0.3ex}{1.41} & \raisebox{0.1ex}{-} & \\[0.2ex]%
\raisebox{0.3ex}{H-CGNR} & \raisebox{0.3ex}{1.57} & \raisebox{0.3ex}{3.54} & \raisebox{0.3ex}{2.21} &
 \raisebox{0.3ex}{2.25\footnote{Ref. \cite{natureche} for C$_{12}$H$_{25}$-CGNR.}} & \raisebox{0.3ex}{-}  & \raisebox{0.3ex}{-} & \raisebox{0.3ex}{-}  & \raisebox{0.3ex}{1.33} & \raisebox{0.3ex}{1.45} & \raisebox{0.3ex}{-} & \\[0.2ex]%
\raisebox{0.3ex}{Cl-CGNR} & \raisebox{0.3ex}{1.45} & \raisebox{0.3ex}{3.39} &
 \raisebox{0.3ex}{2.11} & \raisebox{0.3ex}{2.09\footnote{Ref. \cite{naturecomm}}} & \raisebox{0.3ex}{1.37} & \raisebox{0.3ex}{-} & \raisebox{0.3ex}{-} & \raisebox{0.3ex}{-} & \raisebox{0.3ex}{-} & \raisebox{0.3ex}{1.28} & \\[0.2ex]%
%\raisebox{0.3ex}{C$_{12}$H$_{25}$-CGNR} & \raisebox{0.3ex}{1.57} & \raisebox{0.3ex}{-} &
% \raisebox{0.3ex}{-} & \raisebox{0.3ex}{2.25\footnote{Ref. \cite{naturecomm}}} & \raisebox{0.2ex}{-}  & \raisebox{0.3ex}{-} & \raisebox{0.3ex}{-} & \raisebox{0.3ex}{-} & \raisebox{0.3ex}{-} & \raisebox{0.3ex}{-} & \raisebox{0.3ex}{-} & \\[0.2ex]
\hline\hline

\end{tabular}
%}
\par
%\end{ruledtabular}
\end{center}
%\par
%\par
%\label{tab:PPer}
\end{table*}
%\footnotetext[1]{Ref. \cite{naturecomm}}
%%%%%%%%%%%%%%%%%%%%%%%%%%%%%%%%%%%%%%%%
%%%%%%%%%%%%%FIG4%%%%%%%%%%%%%%%%%%%%%%%%%%%%%%%%%
\begin{figure}[!]
\centering
\includegraphics[width=0.90\columnwidth]{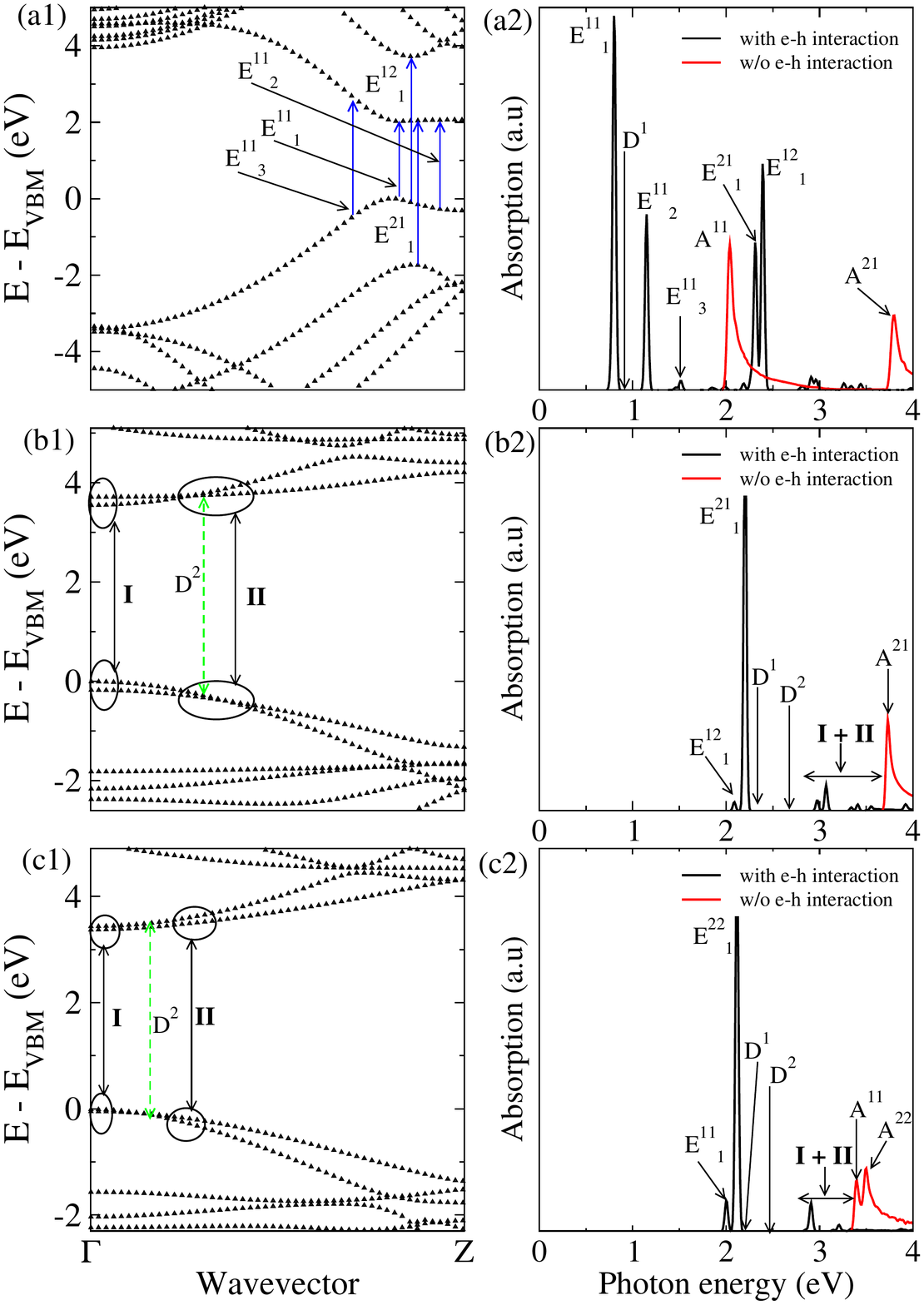}
\caption{\small{Quasiparticle band structure (left)
and optical absorption (right) for a) 4-ZGNR, b) H-CGNR, and c) Cl-CGNR. $A^{ij}$ corresponds to the
absorption at the level of GW-RPA from the $i^{th}$ VB to the $j^{th}$ CB, and $E^{ij}_{S}$ is associated to the corresponding $S^{th}$ exciton
for that transition including e-h interaction.
(a1) The solid (blue)
arrow stands for the wavevector values (or region of values) at which
the excitonic transition occurs. (b1) and (c1) The  green dash-dotted
arrow indicates the position in the k-space at which the dark exciton $D^{2}$ takes place.
The absorption curves were calculated with an artificial gaussian
broadening of 15 meV.
}} \label{fig.4}
%{\rule{4cm}{2cm}}
\end{figure}
%%%%%%%%%%%%%FIG4%%%%%%%%%%%%%%%%%%%%%%%%%%%%%%%%%
The GW-corrected quasiparticle band structures are shown in Fig. \ref{fig.4}.
For the sake of simplicity, the calculations
for 4-ZGNR were performed in the unfolded Brillouin zone. The corrected band structures are very similar to the
PBE ones, except by a a rigid shift of the empty levels. As expected they all display
larger values for the fundamental gap by at least twice the GGA one, as is indicated in Table \ref{table1}.
This increase
over the GGA gap is a signature of the key role that Coulomb interactions
and reduced screening have over low-dimensional systems.\cite{prolouie}

From the
quasi-particle band structures one can obtain the non-interacting electron-hole optical transitions within GW-RPA,
shown as red lines in the right panels of Fig. \ref{fig.4}. These peaks are labeled
as $A^{ij}$, indicating the interband transition from the $i-th$ VB to the $j-th$ CB.

For 4-ZGNR, the non-interacting optical spectrum presents 2 peaks. The first one ($A^{11}$) corresponds to a
continuum of direct interband transitions between the VBM and CBM (ranging from 2.0 to 3.5 eV), which is followed by
another continuum of peaks (starting at 3.8 eV) that correspond to interchanged transitions between the second
 valence band and the first conduction band.\footnote{The $A^{12}$ peak is also present albeit at higher energy.}

For H-CGNR the direct transition from the CBM to the VBM is forbidden. The first observed
transition occurs from VBM to the CBM+1. On the other hand, the non-interacting electron-hole
optical spectrum for Cl-CGNR presents two peaks at $A^{11}$=3.39 and $A^{22}$=3.49 which corresponds
to direct transitions around the $\Gamma$ point. This means that the interchanged conduction bands at the $\Gamma$ point
leads to a different set of transitions. This can be understood from the symmetry of the wavefunctions (Figure S5 of the supplementary
information).

In the presence of electron-hole interactions, bright ($E^{ij}_{S}$) and dark ($D^{S}$) excitonic states arise in the absorption spectrum.
by means of the exciton amplitude of probability, we are able to elucidate the weight associated to each
transition. Based on this weight, we indexed each exciton state peak (black lines in the right panels of Figure \ref{fig.4}) to the
most probable optical transition taking place. The different VBs and CBs involved in those transitions are presented in table S3. We also calculated
the excition binding energies and the results are shown in table \ref{table1}.

The first continuum of transitions for the
pristine GNR gives rise to intense peaks related to bright bound excitonic states located at $E^{11}_{1}$=0.8 eV and
$E^{11}_{2}$=1.15 eV. In addition, resonant excitons ($E^{21}_{1}$ and $E^{12}_{1}$), whose excitation energies lie above the
quasiparticle band gap, also present intense peaks in the absorption spectrum. These resonant states present binding energies higher than the first bound ones
as can be seen in Table \ref{table1}. We also mention that for
energies bellow 1.5 eV, two Dark excitons (not all of them shown) are identified in the absorption spectrum (see
supplementary information). Energywise, each dark exciton arises in the vicinity of the bright ones. For
instance, the first dark exciton, $D^{1}$=0.84 eV, is only 40 meV higher than the bright exciton $E^{11}_{1}$. A similar behavior
has been priviously observed in
8-ZGNR.\cite{zgnlouie}

%In particular, the intense peak at $E^{11}_{1}$, corresponds to the first optical transition from
%the VBM to CBM at a wave vector around four-fifths the $\Gamma$Z path.

%\textcolor{blue}{\st{It should be noticed that within the e-h formalism,
%the number of peaks and their amplitude presents significant differences with respect
%to the non-interacting peaks. % associated to the transitions between VBM and CBM.
%This is a direct consequence of vanishing velocity matrix elements due to the suppression by momentum conservation
%as well as due to the symmetry of the excitonic wavefunctions related to the GNRs (see Supplementary
%information).}}

The e-h optical spectrum for the two CGNR comes from a more complex scenario,
since there are %each peak is constituted by a group of transitions due to
many bands around the Fermi level.
Thus, the excitonic peaks do not
%, contrary to 4-ZGNR where each excitonic peak
belong to a well-defined interband transition. For instance, the
first peak in the H-CGNR absorption spectrum, $E^{12}_{1}=2.09$ eV, and the second peak (the most intense one)
$E^{21}_{1}=2.21$ eV, arise from a group of transitions taking place at the $\Gamma$ point. As previously mentioned they are
indexed following the largest contribution to
the exciton wavefunction, but in fact, they involve combinations of transitions between the two highest valence bands and the two lowest
conduction bands. This includes the originally forbidden direct transitions.

%, labeled as group I in Fig. \ref{fig.4}.
%In addition, two dark exciton states $D^{1}$=2.31 eV and $D^{2}$=2.69 eV, are
%observed, each of them related to group of transitions I and I+II (Figure \ref{fig.4}), respectively. In general,
%dark excitons are responsible for creating channels for nonradiative recombination processes.
% \textcolor{red}{Nevertheless,
%by changing the light polarization direction, it would be possible to optically excited them, which
%might lead to new phenomena.} \todo[inline]{Dangerous statement. The referee might ask why we did not do it}

Accordingly, for Cl-CGNR, in the presence of electron-hole interaction, the exciton states $E^{11}_{1}$=2.0
and $E^{22}_{1}$=2.11 arise from a similar combination involving the two topmost VBs and
the bottommost CBs around the $\Gamma$ point. It should be pointed out
that differently to the H-CGNR case, the presence of Chlorine gives rise to larger weights in the direct optical transition
from the VBM to the CBM.

In both CNRs, there is also a set of less intense peaks in the energy range
from 2.9 - 3.7 eV. These peaks correspond to groups of transitions I and II (shown in the left panel of Figure \ref{fig.4}), which occur at
different wavevectors in two distinct regions, namely close to the $\Gamma$ point and around $k_z=\pi/3a$.
It is worth mentioning that such optical transitions corresponding to groups of
 bands have not been observed before either in armchair GNRs or Chrevon-type GNRs.\cite{qp1,qp2,qp3}
This behavior, however, has been  previously predicted in silicon nanowires\cite{Sinano} and is a
consequence of the proximity between many quasi-particle energy levels at specific wavevectors.

%The dark exciton $D^{1}$=2.18, whose energy is close to the
%excitonic state $E^{22}_{1}$, occurs around the $\Gamma$ point, whereas the dark exciton $D^{2}$=2.45 is located at
%wavevector's values between the $\Gamma$ point and $k_z=\pi/3a$, as shown in Fig.4 (c1) by the
%green dashed arrow.

%\st{which is an important feature sought in materials that can act as donors in photovoltaic devices.}

\begin{figure}[]
\centering
\includegraphics[width=1.0\columnwidth]{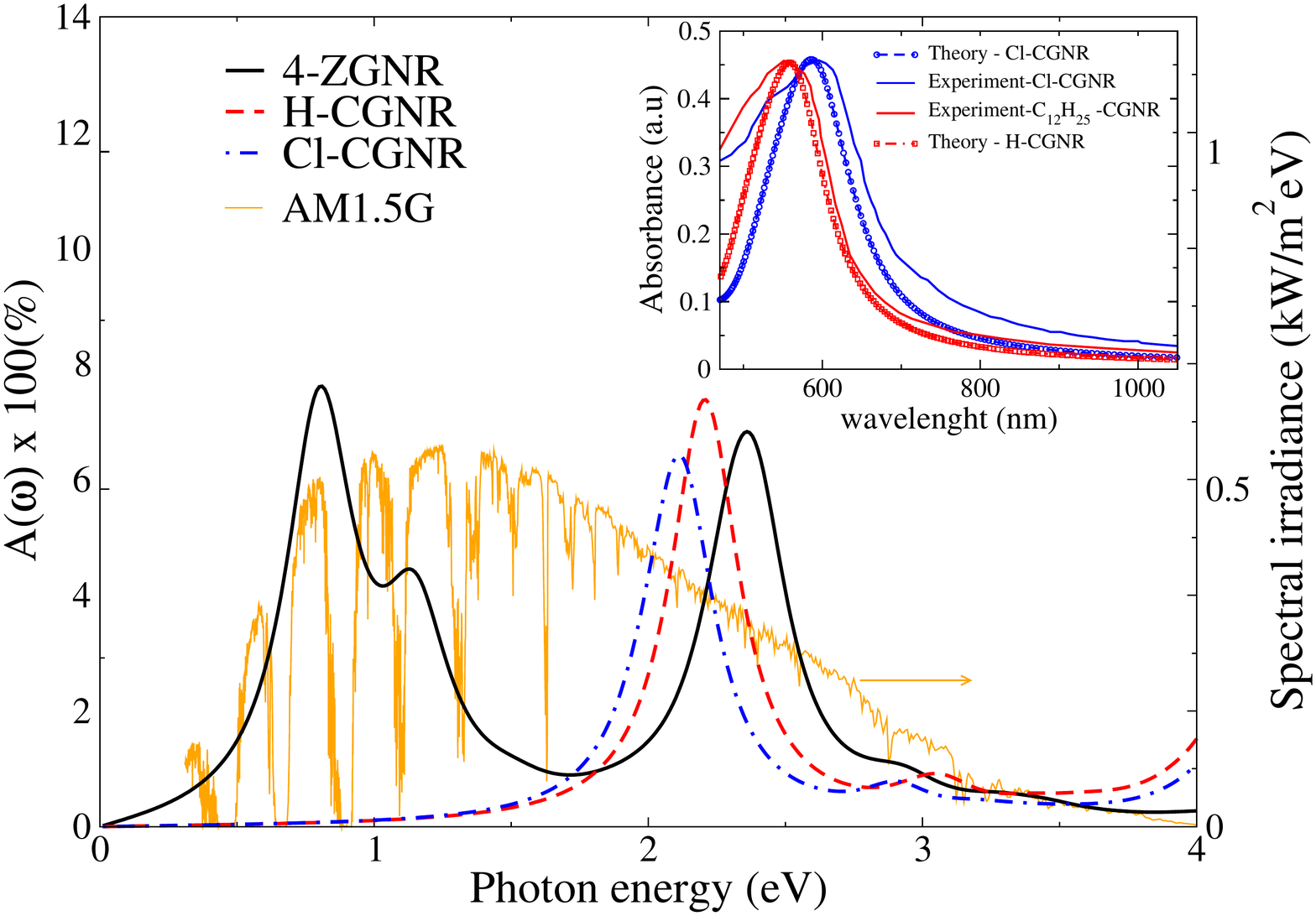}
\caption{\small{Absorbance as a function of
the photon energy for three different GNRs. The  AM1.5 Global irradiance is overlayed to indicate the regions in energy of maximum photon incidence. Inset: Absorbance as a function of photon wavelength
in arbitrary units. The experimental curves were extracted from Ref. \cite{natureche} and \cite{naturecomm}.
For all theoretical results an artificial
Gaussian broadening of 0.15 eV was used.}
} \label{fig.5}
%{\rule{4cm}{2cm}}
\end{figure}

In order to explore the possible use of bottom-up GNRs for photovoltaic applications
the absorbance
\begin{equation}
A(\omega)=1-e^{-\alpha(\omega)\Delta L}
\end{equation}
%\begin{equation}
%A(\omega)=1-e^{\alpha(\omega)\Delta L}
%\end{equation}
was calculated. Here {\small{$\alpha(\omega)=\omega\epsilon_{2}(\omega)/cn$}} is the absorption
coefficient whose values for the GNRs in this work go up to 5$\times$$10^{5}$ $cm^{-1}$
 (see supplementary information). Here $c$, $n$ and $\Delta$L
are the speed of light, refractive index and the dimension of the simulation cell in
the layer-perpendicular direction respectively. The refractive index is assumed to be unity
since the unit cell is constituted mainly by vacuum as described in Ref. \cite{hete}.
In the inset of Fig. \ref{fig.5}, we compare the theoretical e-h absorbance with the experimental
measurements obtained by Narita {\it et al.}\cite{natureche} and Tan {\it et al.}.\cite{naturecomm}
One can note that the position of the peaks are in good agreement with our calculations.
This quantitative agreement between theory and experiment strengthen the crucial
role electron-hole interactions have in predicting accurate optical properties.

%%%%%%%%%%%%%FIG1%%%%%%%%%%%%%%%%%%%%%%%%%%%%%%%%%
%%%%%%%%%%%%%FIG1%%%%%%%%%%%%%%%%%%%%%%%%%%%%%%%%%

Figure \ref{fig.5} depicts the absorbance of the three GNRs as well as the
spectral irradiance of the incident AM1.5 solar spectrum (orange line) for comparison.\cite{am1.5} Both the Cl-CGNR and H-CGNR show
absorbances (related to one peak) ranging from 0.5-7\%. In contrast,
4-ZGNR, presents three intense peaks with absorbance values that go up to 6.5\%
and spreads over the entire visible and near infrared electromagnetic spectrum.
The peaks in the absorbance have a one-to-one correspondence
with the energy of the excitons shown in Figure \ref{fig.4}. We note that the
absorption coefficients have the same order of magnitude as in graphene oxide,\cite{go} and one order
of magnitude higher than in bulk GaAs\cite{gaas} and P3HT\cite{p3ht}, two of the foremost materials
used in photovoltaics due to their high power conversion efficiencies.
Furthermore, these values double the 2.3\% sun light absorption reached for graphene in the visible
spectrum\cite{graf2} and are comparable with the one of monolayer MoS$_{2}$\cite{mos2},
a very promising visible light absorber.

%\todo{Rephrase}
%  e  Although the curves do look similar,
%it should be noted that the differences compared to the experimental values for
%wavelength values located to the right of the peaks might be associated to the solution
%in which the bottom-up GNRs are synthesized\cite{naturecomm} and whose
%effects are not taken into account in our simulations. In fact, similar differences
%in the absorbance measurements performed in solution or in thin films
%have been observed in $C_{60}$ nanorods.\cite{c60}

We then estimate the upper limit for the short-circuit current density related to a donor material in a photovoltaic device,
\begin{equation} \label{curr}
J_{SC}=e\int_{E_{gap}/\hbar}^\infty A(\omega)I_{ph}(\omega)\mathrm{d}\omega,
\end{equation}
where $I_{ph}(\omega)$ is the photon spectral irradiance of the source and $e$ is the electron charge.

The short-circuit current densities for 4-ZGNR, H-CGNR and Cl-CGNR are calculated to be
1.48, 0.54 and 0.51 $mA/cm^{2}$, respectively. These values are already up to 1 order of magnitude larger than nanometer-thick
silicon.\cite{hete}
The value for 4-ZGNR is almost three times that of the H(Cl)-CGNR
since its absorbance covers the energy regions around 1.1 eV.
According to the Shockley-Queisser limit this would provide the most efficient photovoltaic
device.\cite{shoc} Thus, although pristine GNRs have not been realized, they seem to provide an upper limit for the short-circuit current and
absorbance in graphene nanoribbons. Nevertheless, the position of the main peak lies bellow that value and bringing it to higher energies
requires a strategy to further increase the optical gap. This would be hard as the width of the nanoribbon is already at its limit.

The cove-shaped GNRs, on the other hand, already give significant improvement over graphene. The first exciton also has a higher
binding energy compared to the pristine case, but most importantly its position lies in the high energy end of the solar spectrum.
Consequently one could use wider molecular precursors to decrease the optical gap
and bring the transitions closer to the energy where efficiency is a maximum.

%Furthermore, one can also estimate the $J_{SC}$ for nanometer thick
%silicon,\cite{hete} which is about 0.1 $mA/cm^{2}$, compared to
%$J_{SC}$ in the GNRs. This is , which reinforces the potential of GNRs for acting
%as donor materials in photovoltaic.

% Nevertheless, cove-shaped ribbons already
%At the same time, these CGNRs  This might lead to
%longer diffusion lengths. Furthermore, there are a number of strategies that can be adopted to improve
%the $J_{SC}$. For instance, by stacking two or three different monolayers of NRs,
%one should expect % \st{the reduction of the optical band gap as well as the}
%a considerable increase in the absorbance, %\st{(double or even triple)},
%which is an important factor for improving the $J_{SC}$ values and the power conversion efficiency of a photovoltaic
%device. %\st{Certainly, further work exploring the strategies mentioned above and a correct
%choice for a acceptor material are required to establish without any doubt the relevance
%of bottom-up GNRs in photovoltaic.}

%\todo[inline]{I should discuss that while ZGNR have the highest absorbance, the CGNR give a better perspective as the exciton energy could be
%made smaller by increasing the width of the precursor molecule}
%\todo[inline]{Although the systems seem similar, they present transitions with different features. I did not
%understand this ???}

\section{Conclusions}

We have carried out DFT calculations combined with many-body perturbation theory
to study the electronic and optical properties of novel atomically well-defined
graphene nanoribbons. %The electronic structure shown that the presence of
%dodecane chains bound at the edges of the benzo-fused rings do not play
%a significant role over the states surrounding the Fermi level.
Through the comparison with experimental measurements, we were able to show
that an accurate description of the GNR's optical properties
involves the inclusion of electron-hole correlation effects.

We also found that the excitonic peaks in the absorption spectrum of cove-shaped graphene nanoribbons are
consequence of a group of transitions involving the first and second conduction and valence
bands. %We also calculated the absorption coefficient for the GNRs which turn out to be
%one order of magnitude higher than those found for GaAs and P3HT.
We also note that different functionalization of the edges can lead to changes in the character of the band
transitions.

Finally, we estimated the short-circuit current density for the nanometer thick bottom-up GNRs,
finding attractive values that go up to one order of magnitude higher than those
estimated for a nanometer-thick Si and GaAs, foremost materials used in photovoltaic.
These interesting results show the possibility of use bottom-up GNRs as promising donor materials
in photovoltaics. In particular, we propose that cove-shaped GNRs could be assembled using wider precursor molecules to tailor the optical gap, thus enhancing even further their potential for photovoltaics

\section{Acknowledgements}

The authors acknowledge the financial support from the Brazilian agency FAPESP. We also thank M. Menezes for fruitfull discussions. The calculations were
carried at GRID-UNESP and CENAPAD/SP.

\bibliographystyle{unsrtnat}
%\bibliography{ref_v2.bib}

\begin{thebibliography}{54}
\providecommand{\natexlab}[1]{#1}
\providecommand{\url}[1]{\texttt{#1}}
\expandafter\ifx\csname urlstyle\endcsname\relax
  \providecommand{\doi}[1]{doi: #1}\else
  \providecommand{\doi}{doi: \begingroup \urlstyle{rm}\Url}\fi

\bibitem[Neto et~al.((2009))Neto, Guinea, Peres, Novoselov, and Geim]{rev}
Castro Neto, A.~H., Guinea, F., Peres, N.~M.~R., Novoselov, K.~S. \& Geim, A.~K.
\newblock The electronic properties of graphene.
\newblock \emph{Rev. Mod. Phys.} \textbf{81}, 109 (2009).

\bibitem[Sarma et~al.(2011)Sarma, Adam, Hwang, and Rossi]{rev2}
Das Sarma, S., Adam, S., Hwang, E.~H. \& Rossi, E.
\newblock Electronic transport in two-dimensional graphene.
\newblock \emph{Rev. Mod. Phys.} \textbf{83}, 407 (2011).

\bibitem[Novoselov et~al.(2012)Novoselov, Fal'ko, Colombo, Gellert, Schwab, and
  Kim.]{grap}
Novoselov, K.~S. {\it et al}.
\newblock A roadmap for graphene.
\newblock \emph{Nature} \textbf{490}, 192 (2012).

\bibitem[Avouris and Dimitrakopoulos(2012)]{prospec1}
Avouris, P. \& Dimitrakopoulos, C.
\newblock Graphene: synthesis and applications.
\newblock \emph{Materials Today}, \textbf{15}, 86 (2012).

\bibitem[Geim(2009)]{prospec}
Geim, A.~K.
\newblock Graphene: Status and prospects.
\newblock \emph{Science} \textbf{324}, 1530 (2009).

\bibitem[Bonaccorso et~al.(2010)Bonaccorso, Sun, Hasan, and Ferrari]{gopto}
Bonaccorso, F., Sun, Z., Hasan, T. \& Ferrari., A.~C.
\newblock Graphene photonics and optoelectronics.
\newblock \emph{Nature Photonics} \textbf{4}, 611 (2010).

\bibitem[Wang et~al.(2014)Wang, Ball, Barea, Abate, Alexander-Webber, Huang,
  Saliba, Mora-Sero, Bisquert, Snaith, and Nicholas]{gphoto}
Wang, J.~T-W. {\it et al}.
\newblock Low-temperature processed electron collection layers of graphene/TiO2
  nanocomposites in thin film perovskite solar cells.
\newblock \emph{Nano Lett.}, \textbf{14}, 724 (2014).

\bibitem[Bernardi et~al.(2013)Bernardi, Palummo, and Grossman]{hete}
Bernardi, M., Palummo, M. \& Grossman, J.~C. 
\newblock Extraordinary sunlight absorption and one nanometer thick
  photovoltaics using two-dimensional monolayer materials.
\newblock \emph{Nano Lett.}, \textbf{13}, 3664 (2013).

\bibitem[Su et~al.(2012)Su, Lan, and Wei]{frenk_photo}
Su, Y.-W., Lan, S.-C. \& Wei, K.-H.
\newblock Organic photovoltaics.
\newblock \emph{Materials Today}, \textbf{15}, 554 (2012).

\bibitem[Peyghan et~al.(2013)Peyghan, Noei, and Tabar]{adsorp}
Peyghan, A.~A., Noei, M. and Tabar, M.~B.
\newblock A large gap opening of graphene induced by the adsorption of Co on
  the Al-doped site.
\newblock \emph{Jour. Mol. Mod.}, \textbf{19}, 3007 (2013).

\bibitem[Ulstrup et~al.(2013)Ulstrup, Nilsson, Miwa, Balog, Bianchi, r, and
  Hofmann]{adsorp2}
Ulstrup, S. {\it et al}.
\newblock Electronic structure of graphene on a reconstructed Pt(100) surface:
  Hydrogen adsorption, doping, and band gaps.
\newblock \emph{Phys. Rev. B} \textbf{88}, 125425 (2013).

\bibitem[Ni et~al.(2008)Ni, Yu, Lu, Wang, Feng, and Shen]{strain}
Ni, Z. {\it et al}.
\newblock Uniaxial strain on graphene: Raman spectroscopy study and band-gap
  opening.
\newblock \emph{ACS Nano} \textbf{2}, 2301 (2008).

\bibitem[Choi et~al.(2010)Choi, Jhi, and Son]{strain2}
Choi, S-M., Jhi, S.-H. \& Son, Y.-W.
\newblock Effects of strain on electronic properties of graphene.
\newblock \emph{Phys. Rev. B} \textbf{81}, 081407 (2010).

\bibitem[Guo et~al.(2010)Guo, Liu, Chen, Zhu, Fang, and Gong]{doped}
Guo, B. {\it et al}.
\newblock Controllable n-doping of graphene.
\newblock \emph{Nano Lett.} \textbf{10},4975 (2010).

\bibitem[Ci et~al.(2010)Ci, Song, Jin, Jariwala, Wu, Li, Srivastava, Wang,
  Storr, Balicas, Liu, and Ajayan]{doped2}
Ci, L. {\it et al}.
\newblock Atomic layers of hybridized boron nitride and graphene domains.
\newblock \emph{Nature Mat.} \textbf{9}, 430 (2010).

\bibitem[Son et~al.(2006)Son, Cohen, and Louie]{ribbon}
Son, Y-W., Cohen,  M.~L. \& Louie, S.~G. 
\newblock Energy gaps in graphene nanoribbons.
\newblock \emph{Phys. Rev. Lett.} \textbf{97}, 216803 (2006).

\bibitem[Han et~al.(2007)Han, Ozyilmaz, Zhang, and Kim]{ribbon1}
Han, M.~Y., Ozyilmaz,  B., Zhang, Y. \& Kim, P.
\newblock Energy band-gap engineering of graphene nanoribbons.
\newblock \emph{Phys. Rev. Lett.} \textbf{98}, 206805 (2007).

\bibitem[Singh and Yakobson(2009)]{nanorods}
Singh, A.~K. \& Yakobson, B.~I.
\newblock Electronics and magnetism of patterned graphene nanoroads.
\newblock \emph{Nano Lett.} \textbf{9}, 1540 (2009).

\bibitem[de~Almeida et~al.(2013)de~Almeida, abd A.~K~Singh, Fazzio, and
  da~Silva]{nanoroadsJames}
De~Almeida, J.~M., Rocha, A.~R., K~Singh, A., Fazzio, A. \& Da~Silva, A.~J.~R.
\newblock Electronic transport in patterned graphene nanoroads.
\newblock \emph{Nanotechnology} \textbf{24}, 495201 (2013).

\bibitem[Li et~al.(2008)Li, Wang, Zhang, Lee, and Dai]{ribbon2}
Li, X., Wang, X., Zhang, L., Lee, S., \& Dai, H.
\newblock Chemically derived, ultrasmooth graphene nanoribbon semiconductors.
\newblock \emph{Science} \textbf{319}, 1229 (2008).

\bibitem[Ritter and Lyding(2009)]{ribbon3}
Ritter, K.~A. \& Lyding, J.~W.
\newblock The influence of edge structure on the electronic properties of
  graphene quantum dots and nanoribbons.
\newblock \emph{Nature Mat.} \textbf{8}, 235 (2009).

\bibitem[Osella et~al.(2012)Osella, Narita, Schwab, Hernandez, Feng, M\"ullen,
  and Beljonne]{opto}
Osella, S. {\it et al}.
\newblock Graphene nanoribbons as low band gap donor materials for organic
  photovoltaics: Quantum chemical aided design.
\newblock \emph{ACS Nano} \textbf{6}, 5539 (2012).

\bibitem[Yang et~al.(2007{\natexlab{a}})Yang, Cohen, and Louie.]{qp2}
Yang, L., Cohen, M.~L. \& Louie, S.~G.
\newblock Excitonic effects in the optical spectra of graphene nanoribbons.
\newblock \emph{Nano Lett.} \textbf{7}, 3112 (2007).

\bibitem[Bai et~al.(2009)Bai, Duan, and Huang]{lito}
Bai, J., Duan, X. \& Huang, Y.
\newblock Rational fabrication of graphene nanoribbons using a nanowire etch
  mask.
\newblock \emph{Nano Lett.} \textbf{9}, 2083 (2009).

\bibitem[Kosynkin et~al.(2009)Kosynkin, Higginbotham, Sinitskii, Lomeda,
  Dimiev, Price, and Tour]{unzip}
Kosynkin, D.~V. {\it et al}.
\newblock Longitudinal unzipping of carbon nanotubes to form graphene
  nanoribbons.
\newblock \emph{Nature} \textbf{458}, 872 (2009).

\bibitem[Wu et~al.(2010)Wu, Ren, Gao, Liu, Zhao, and Cheng]{soni}
Wu, Z-S. {\it et al}.
\newblock Efficient synthesis of graphene nanoribbons sonochemically cut from
  graphene sheets.
\newblock \emph{Nano Res.} \textbf{3}, 16 (2010).

\bibitem[Genorio and Znidarsic(2014)]{bottom-up}
Genorio, B. \& Znidarsic, A.
\newblock Functionalization of graphene nanoribbons.
\newblock \emph{J. Phys. D: Appl. Phys.} \textbf{47}, 094012 (2014).

\bibitem[Li et~al.(2010)Li, Gao, Motta, Negri, and Wang]{solu}
Li, Y., Gao, J., Di Motta, S., Negri, F. \& Wang, Z.~J.
\newblock Tri-n-annulated hexarylene: An approach to well-defined graphene
  nanoribbons with large dipoles.
\newblock \emph{Am. Chem. Soc} \textbf{132}, 4208 (2010).

\bibitem[Tan et~al.(2013)Tan, Yang, Parvez, Narita, Osella, Beljonne, Feng, and
  M\"ullen]{naturecomm}
Tan, Y-Z. {\it et al}.
\newblock Atomically precise edge chlorination of nanographenes and its
  application in graphene nanoribbons.
\newblock \emph{Nature Commun.} \textbf{4}, 2646 (2013).

\bibitem[Narita et~al.(2014)Narita, Feng, Hernandez, Jensen, Bonn, Yang,
  Verzhbitskiy, Casiraghi, Hansen, Koch, Fytas, Ivasenko, Li, Mali, Balandina,
  Mahesh, Feyter, and M\"ullen]{natureche}
Narita, A. {\it et al}.
\newblock Synthesis of structurally well-defined and liquid-phase-processable
  graphene nanoribbons.
\newblock \emph{Nature Chem.} \textbf{6}, 126 (2014).

\bibitem[Vo et~al.(2014)Vo, Shekhirev, Kunkel, Morton, Berglund, Kong, Wilson,
  Dowben, Enders, and Sinitskii]{naturecomm2}
Vo, T.~H. {\it et al}.
\newblock Large-scale solution synthesis of narrow graphene nanoribbons.
\newblock \emph{Nature Commun.} \textbf{5}, 3189 (2014).

\bibitem[Cai et~al.(2010)Cai, Ruffieux, Jaafar, Bieri, Braun, Blankenburg,
  Muoth, Seitsonen, Saleh, Feng, M\"ullen, and Fasel]{cai}
Cai, J. {\it et al}.
\newblock Atomically precise bottom-up fabrication of graphene nanoribbons.
\newblock \emph{Nature} \textbf{466}, 470 (2010).

\bibitem[Yang et~al.(2007{\natexlab{b}})Yang, Park, Son, Cohen, and
  Louie]{loui}
Yang, L., Park, C-H., Son, Y-W., Cohen, M.~L. \& Louie, S.~G.
\newblock Quasiparticle energies and band gaps in graphene nanoribbons.
\newblock \emph{Phys. Rev. Lett.} \textbf{99}, 186801 (2007){\natexlab{b}}.

\bibitem[Hohenberg and Kohn(1964)]{dft1}
Hohenberg, P. \& Kohn, W.
\newblock Inhomogeneous electron gas.
\newblock \emph{Phys. Rev.} \textbf{136}, B864 (1964).

\bibitem[Kohn and Sham(1965)]{dft2}
Kohn, W. \& Sham, L.~J.
\newblock Self-consistent equations including exchange and correlation effects.
\newblock \emph{Phys. Rev.} textbf{140}, A1133 (1965).

\bibitem[Perdew et~al.(1996)Perdew, Burke, and Ernzerhof]{pbe}
Perdew, J.~P., Burke, K. \& Ernzerhof, M.
\newblock Generalized gradient approximation made simple.
\newblock \emph{Phys. Rev. Lett.} \textbf{77}, 3865 (1996).

\bibitem[et~al.(2009)]{QE}
Giannozzi, P. {\it et al}.
\newblock Quantum espresso: a modular and open-source software project for
  quantum simulations of materials.
\newblock \emph{J. Phys. Condens. Matter} \textbf{21}, 395502 (2009).

\bibitem[Rohlfing and Louie(2000)]{bse}
Rohlfing, M. \& Louie, S.~G.
\newblock Electron-hole excitations and optical spectra from first principles.
\newblock \emph{Phys. Rev. B} \textbf{62}, 4927 (2000).

\bibitem[Fetter and Walecka(1971)]{tammdancoff}
Fetter, A. \& Walecka, J.~D.
\newblock \emph{Quantum Theory of Many Particle Systems}.
\newblock McGraw-Hill Book Company: San Francisco, 1971.
\newblock pp. 538 $-$ 539.

\bibitem[Marinopoulos et~al.(2003)Marinopoulos, Reining, Rubio, and Vast]{23}
Marinopoulos, A.~G., Reining, L., Rubio, A. \& Vast, N.
\newblock Optical and loss spectra of carbon nanotubes: Depolarization effects
  and intertube interactions.
\newblock \emph{Phys. Rev. Lett.} \textbf{91}, 046402 (2003).

\bibitem[Deslippe et~al.(2012)Deslippe, Samsonidze, Strubbe, Jain, Cohen, and
  Louie]{bgw}
Deslippe, J. {\it et al}.
\newblock \emph{Comput. Phys. Commun.} \textbf{183}, 1269 (2012).

\bibitem[Jiang et~al.(2007)Jiang, Sumpter, and Dai]{afmzgnr}
Jiang, D.-E., Sumpter, B.~G. \& Dai, S.
\newblock Unique chemical reactivity of a graphene nanoribbon’s zigzag edge.
\newblock \emph{Jour. of Chem. Phys.} \textbf{126}, 134701 (2007).

\bibitem[Deslippe and Louie(2008)]{prolouie}
Deslippe, J. \& Louie, S.~G.
\newblock Excitons and many-electron effects in the optical response of carbon
  nanotubes and other one-dimensional nanostructures.
\newblock \emph{Proc. SPIE} \textbf{6892}, 68920U--1 (2008).

\bibitem[Yang et~al.(2008)Yang, Cohen, and Louie]{zgnlouie}
Yang, L., Cohen, M.~L. \& Louie, S.~G.
\newblock Magnetic edge-state excitons in zigzag graphene nanoribbons.
\newblock \emph{Phys. Rev. Lett.} \textbf{101}, 186401 (2008).

\bibitem[Prezzi et~al.(2008)Prezzi, Varsano, Ruini, Marini, and Molinari]{qp1}
Prezzi, D., Varsano,  D., Ruini,  A., Marini, A. \& Molinari, E.
\newblock Optical properties of graphene nanoribbons: The role of many-body
  effects.
\newblock \emph{Phys. Rev. B} \textbf{77}, 041404 (2008).

\bibitem[Wang and Wang.(2012)]{qp3}
Wang, S. \& Wang, J.
\newblock Quasiparticle energies and optical excitations in chevron-type
  graphene nanoribbon.
\newblock \emph{Jour. Phys. Chem. C.} \textbf{116}, 10193 (2012).

\bibitem[Yang et~al.(2007{\natexlab{c}})Yang, Spataru, Louie, and Chou]{Sinano}
Yang, L., Spataru, C.~D., Louie, S.~G. \& Chou, M.~Y.
\newblock Enhanced electron-hole interaction and optical absorption in a
  silicon nanowire.
\newblock \emph{Phys. Rev. B.} \textbf{75}, 201304(R), (2007){\natexlab{c}}.

\bibitem[am1()]{am1.5}
\emph{http://rredc.nrel.gov/solar/spectra/am1.5/}.

\bibitem[Sokolov et~al.(2014)Sokolov, Morozov, McDonald, Vietmeyer, Hodak, and
  Kuno.]{go}
Sokolov, D.~A. {\it et al}.
\newblock Direct observation of single layer graphene oxide reduction through
  spatially resolved, single sheet absorption/emission microscopy.
\newblock \emph{Nano Lett.} \textbf{14}, 3172 (2014).

\bibitem[Palik(1998)]{gaas}
E.~D. Palik.
\newblock \emph{Handbook of Optical Constants of Solids}, volume~3.
\newblock Academic Press: New York, 1998.

\bibitem[Cook et~al.(2008)Cook, Furubea, and Katoh]{p3ht}
Cook, S., Furubea, A. \& Katoh, R.
\newblock Analysis of the excited states of regioregular polythiophene p3ht.
\newblock \emph{Energy Environ. Sci.} \textbf{1}, 294 (2008).

\bibitem[Nair et~al.(2008)Nair, Blake, Grigorenko, Novoselov, Booth, Stauber,
  Peres, and Geim]{graf2}
Nair, R.~R. {\it et al}.
\newblock Fine structure constant defines visual transparency of graphene.
\newblock \emph{Science} \textbf{320}, 5881 (2008).

\bibitem[Mak et~al.(2010)Mak, Lee, Hone, Shan, and Heinz]{mos2}
Mak, K.~F., Lee,  C., Hone, J., Shan, A. \& Heinz, T.~F.
\newblock Atomically thin MoS2: A new direct-gap semiconductor.
\newblock \emph{Phys. Rev. Lett.} \textbf{105}, 136805 (2010).

\bibitem[Shockley and Queisser(1961)]{shoc}
Shockley, S. \& Queisser, H.~J.
\newblock Detailed balance limit of efficiency of p‐n junction solar cells.
\newblock \emph{J. Appl. Phys.} \textbf{32}, 510 (1961).

\end{thebibliography}

\end{document}